# Eigenmodes of twisted spin-waves in a thick ferromagnetic nanodisk

Peiyuan Huang[1,2] and Ruifang Wang[1,2*]

[1]Department of physics, Xiamen University, Xiamen 361005, China

[2]Key Laboratory of Low Dimensional Condensed Matter Physics, Department of Education of Fujian Province, Xiamen University, Xiamen 361005, China

Email: wangrf@xmu.edu.cn

## Abstract

Magnetic vortex is topologically nontrivial and commonly found in ferromagnetic nanodisks. So far, three classes spin-wave eigenmodes, i.e., gyrotropic, azimuthal and radial modes, have been identified in ferromagnetic nanodisks. Here, using micromagnetic simulation and analytical calculation, we reveal twisted spin-wave modes in a thick permalloy ($Ni_{0.8}Fe_{0.2}$) nanodisk. The twisted spin-waves carry topological charges, which sign depends on the core polarity of the magnetic vortex in the nanodisk. By applying rotating magnetic fields at one end of the sample, we observe continuous generation of twisted spin-waves that have characteristic spiral phase front and carry topological charge $l = \pm 1$ and $\pm 2$. The dispersion relation of twisted spin-waves is derived analytically and the result is in good agreement with micromagnetic numerical calculations.

# 1. Introduction

In ferromagnetic nanodisks, the competition between exchange and magnetostatic interactions favors a magnetic vortex (MV) state, in which the magnetization circulates in-plane around a core where the magnetization turns perpendicular to the disk plane[1,2]. The topologically nontrivial MV state is featured with a vortex core (VC) having upward or downward magnetization, which defines the core polarity $p = +1$ or $-1$. The counterclockwise or clockwise circulation of magnetization, on the other hand, determines the vortex chirality $c = +1$ or $-1$. The spin dynamics and spin-wave eigenmodes in small magnetic structures are of great interest to the study of magnonics[3,4], and applications such as logic circuits[5,6] and high-density data storage devices[5,7].

To date, three classes of spin-wave eigenmodes, namely the gyrotropic[7,8], azimuthal[9-12] and radial[9,10,13], have been identified in MV states. It is noteworthy that the topology of azimuthal spin-wave is similar to twisted waves, which carry orbital angular momentum and have been discovered in photon[14,15], electron[16,17], neutron[18,19] and phonon[20,21] beams in the past three decades, and most recently in magnons[22-25]. The characteristic phase front of twisted beams spirals around its symmetric axis and can be expressed as $\exp(-il\phi)$[14]. The azimuthal angle $\phi$ is defined in the beam's cross section, and $l$ is the topological charge, which can be any integer. The topological charge $l$ can be far bigger than one, so the twisted waves can be utilized in multiplex communication and quantum information technologies[16,26]. Interestingly, the azimuthal spin-wave modes in a MV propagate around the core, which is nearly standstill. The oscillation phase of azimuthal spin-wave, similar to vortex wave, depends on the azimuthal angle as $\exp(-im\phi)$, where $m$ denotes the azimuthal mode number, with $|2m|$ nodes in the azimuthal dimension[12]. However, unlike the vortex waves, which have spiral phase front, the azimuthal spin-waves circulate in the disk plane and their wave front is planar.

Here, by conducting micromagnetic simulations and analytical calculations, we reveal twisted spin-wave modes in a ferromagnetic nanodisk with thickness comparable to its diameter. Twisted spin-wave eigenmodes carrying topological charge $l = \pm 1, \pm 2$ are observed, after exerting a pulsed magnetic field at one end of the sample. Continuous generation of twisted spin-waves is demonstrated by applying rotating magnetic fields applied at one end of the sample. We manifest that the twisted spin-waves propagate not only azimuthally but axially, which means that the phase front spirals around the disk axis. We derive the dispersion relation of twisted spin-waves analytically and the result is in good agreement with the micromagnetic numerical calculations.

## 2. Micromagnetic simulation model

Figure 1(a) illustrates our model system, a 200 nm thick permalloy (Py: $Ni_{0.8}Fe_{0.2}$) disk with a diameter of 200 nm as well. The sample is initially in a stable MV state with $c = +1$ and $p = +1$.[7,13] Typical material parameters for Py are used in the micromagnetic simulations[27]: saturation magnetization $M_S = 800 \ KA/m$, exchange stiffness constant $A_{ex} = 13 \ pJ/m$, Gilbert damping constant $\alpha = 0.01$ and zero magnetocrystalline anisotropy. The mesh cell size is $2 \times 2 \times 5$ nm³.

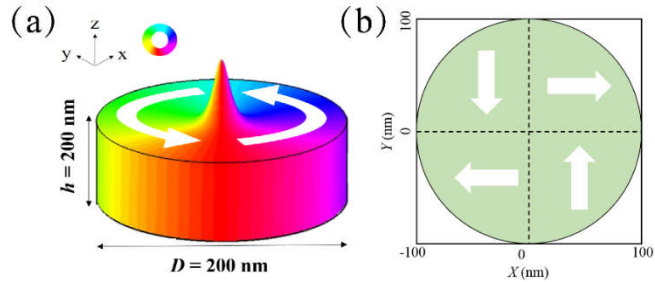

**Fig. 1.** (a) Schematic view of the cylindrical permalloy sample with radius of 100 nm and height of 200 nm. The sample is in a MV state having polarity $p = +1$ and chirality $c = +1$. The colors illustrate the in-plane component of magnetizations, as indicated by the color wheel. (b) Schematic view of the position dependent excitation field, applied at the bottom layer of the sample, to excite the twisted spin-waves with topological charge $l = \pm 2$. White arrows denote that the field vector $\vec{e_p}$ orients along $\vec{e}_x$, $-\vec{e}_y$, $-\vec{e}_x$ and $\vec{e}_y$, at $t$ = 0 ps, in the 1st to 4th quadrants, respectively.

## 3. Eigenmodes of twisted spin-waves

To stimulate spin-wave oscillation, an in-plane sinc-function field, $\vec{B}_{sinc-1}(t) = \vec{e}_x B_{sinc} = \vec{e}_x B_0 \sin[2\pi f_0(t-t_0)]/[2\pi f_0(t-t_0)]$, with $B_0 = 1$ mT, $t_0 = 1$ ns and cut-off frequency $f_0 = 20$ GHz, is applied at the bottom layer of the sample ($0 \leq z \leq 5$ nm). The temporal oscillation of $\langle m_x \rangle = \langle M_x \rangle / M_s$, the x-component of magnetization averaged over the entire disk, is shown in the inset of Fig 2(a).[10,28] Subsequent fast Fourier transformation (FFT) on $\langle m_x \rangle$ creates the FFT power spectrum[10] illustrated in Fig.2(a). The resonance peak at $f$ = 0.35 GHz corresponds to the gyrotropic mode[7,8] of the MV. In addition, eight peaks are found at $f$ = 1.6, 3.5, 4.8, 5.9, 7.3, 8.1, 8.4 and 9.4 GHz. After conducting inverse fast Fourier transformation (IFFT)[10] at these eigenfrequencies, we obtain the FFT amplitude distributions of the corresponding eigenmodes. Figure 3(a) demonstrates the FFT amplitude distribution images of the eigenmode with eigenfrequency $f$ = 3.5 GHZ. It is seen that the wave front rotates clockwise while the wave propagates along z and there are two nodes in the azimuthal dimension, indicating that the mode carries topological charge $l = -1$[14,22,23]. Results on the other seven eigenmodes are similar and not shown here.

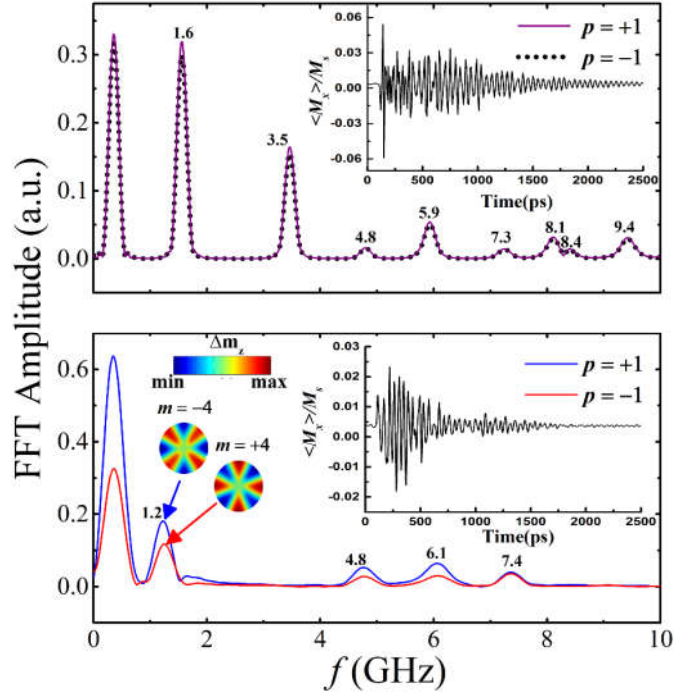

Fig. 2. FFT power spectrum of $<m_x>$ after the sinc-function field $\vec{B}_{sinc-1}(t)$ (a) and $\vec{B}_{sinc-2}(t)$ (b) is exerted on the bottom layer of the sample. The temporal oscillations of $<m_x>$ after the application of external field is plotted in the insets. (a) The two FFT power spectrums of the sample, when the initial polarity is set as $p = +1$ or $p = -1$, are the same. The excitation field $\vec{B}_{sinc-1}(t)$ stimulates twisted spin-wave eigenmodes carrying topological charge $l = -1$ or $l = +1$, at frequencies of 1.6, 3.5, 4.8, 5.9, 7.3, 8.1, 8.4 and 9.4 GHz. (b) The amplitude of the FFT power spectrum for initial vortex state with polarity $p = +1$ is larger than that of the initial state with $p = -1$. Both FFT power spectrums show twisted spin-wave eigenmodes carrying topological charge $l = -2$ or $l = +2$, at frequencies of 4.2, 4.8, 6.1 and 8.4 GHz. The resonance peak at $f$ = 1.2 GHz corresponds to azimuthal spin-wave mode with mode number $m = -4$ and $m = +4$ for initial state with $p = +1$ and $p = -1$, respectively, as shown in the inset of (b).

To excite twisted spin-waves with greater topological charges, a position dependent sinc-function field $\vec{B}_{sinc-2}(t) = \vec{e_p} B_{sinc}$, schematically shown in Fig. 1(b), is exerted on the bottom layer of the nanodisk. After conducting FFT analysis on the temporal variation of $\langle m_x \rangle$, we obtain the FFT power spectrum shown in Fig. 2(b). The resonance peaks at $f$ = 4.8, 6.1, and 7.4 GHz correspond to twisted spin-wave modes with topological charge $l = -2$. FFT amplitude distribution images of the $f = 7.4$ GHz mode, at different layers of the disk, is shown in Fig. 3(c). The eigenfrequency at $f = 1.2$ GHz, on the other hand, corresponds to an azimuthal spin-wave with mode number[12] $m = +4$ for the MV state with $p = -1$ and $m = -4$ for the MV with $p = +1$, respectively (see the inset of Fig. 2(b)). The eigenfrequency of this azimuthal mode is surprisingly low, considering that azimuthal spin-waves previously found in thin nanodisks are on the order of 10 GHz[11,12].

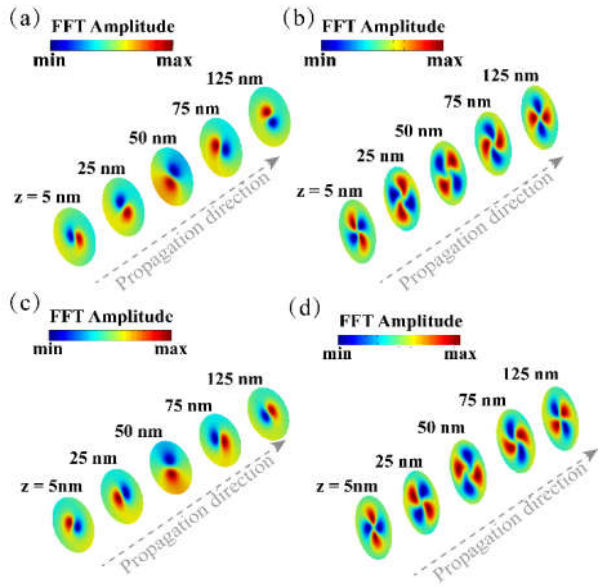

**Fig. 3.** FFT amplitude distribution images of twisted spin-wave modes at different layers of the nanodisk. Images in (a) and (c) are obtained after application of the excitation fields $\vec{B}_{sinc-1}(t)$, and the eigenfrequency is $f = 3.5$ GHz. The vortex modes carry topological charge $l = -1$ [(a)] and $l = +1$ [(c)], when the sample is in a MV state having polarity $p = +1$ and $p = -1$ respectively. Images in (b) and (d) are obtained after application of the excitation fields $\vec{B}_{sinc-2}(t)$, and the eigenfrequency is $f = 7.4$ GHz. The vortex modes carry topological charge $l = -2$ [(b)] and $l = +2$ [(d)], when the sample is in a MV state having polarity $p = +1$ and $p = -1$ respectively.

It is noteworthy that, when the MV state has chirality $c = +1$ and core polarity $p = +1$, the twisted spin-wave modes have negative topological charges, i.e., $l = -1$ or $l = -2$. To investigate the dependence of topological charge carried by twisted spin-wave on the core polarity, the initial MV state is reset to have $c = +1$ and $p = -1$. And additional micromagnetic simulations are carried out. The obtained FFT power spectrums of the temporal oscillation of $\langle m_x \rangle$, after application of excitation field $\vec{B}_{sinc-1}(t)$, coincides with the result of the MV state having $c = +1$ and $p = +1$, as shown in Fig 2(a). Now the corresponding twisted spin-wave modes have topological number $l = +1$. The FFT amplitude distribution images of the eigenmode with eigenfrequency $f$ = 3.5 GHz is illustrated in Fig. 3(c), which shows the wave front rotates counterclockwise while the wave propagates along z. When the field $\vec{B}_{sinc-2}(t)$ is applied, the FFT power spectrum obtained is similar to the result of the sample having $c = +1$ and $p = +1$. The only difference is in the height of resonance peaks, as shown in Fig. 2(b). Now the corresponding twisted spin-wave modes carry topological charge $l = +2$. FFT amplitude distribution images of the eigenmode with eigenfrequency $f$ = 7.4 GHz is shown in Fig. 3(d).

4. **Continuous excitation of twisted spin-waves**

After revealing the twisted spin-wave modes, we demonstrate continuous excitation of these spin-waves in the nanodisk. A rotating magnetic field, with amplitude of 5 mT and tuned to the eigenfrequency of 3.5 GHz, is applied at the bottom layer of the sample. For the sample initially in a MV state with $p = +1$ and $c = +1$, the excitation field rotates counterclockwise in the x-y plane. Fig. 4(a) illustrates snapshots of the twisted spin-waves, 300 ps after application of the rotating field. It is seen that the wave front rotates clockwise indicating topological charges $l = -1$. The spiral phase front of the vortex wave is shown in Fig. 4(c). To generate vortex wave with $l = +1$, the initial MV state is set to have $p = -1$ and $c = +1$ and a clockwise rotating field is applied to the sample. Snap shots of the twisted spin-wave and its spiral phase front is shown in Figs. 4(b) and 4(d), respectively.

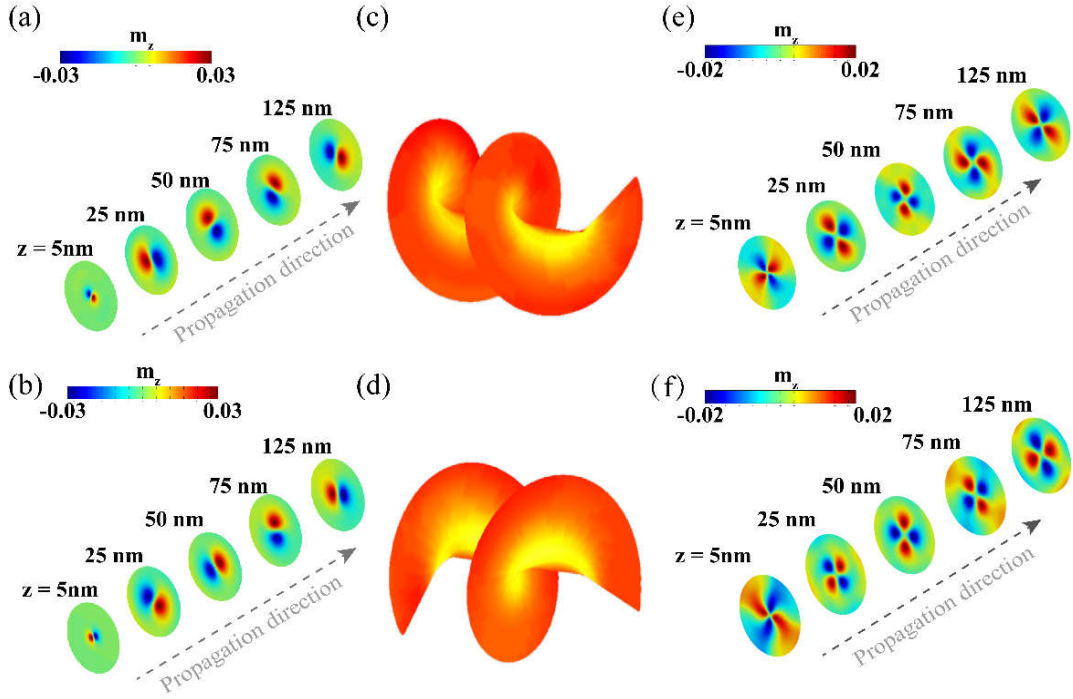

Fig. 4. Snapshots of the z-component of the twisted spin-waves with (a) $l=-1$ and (b) $l=+1$, 300 ps after application of a 3.5GHz field rotating clockwise and clockwise, respectively. The corresponding phase fronts of the vortex waves are illustrated in (c) and (d), respectively. (e) and (f) illustrates the twisted spin-waves with $l=-2$ and $l=+2$, 300 ps after application of a 7.0 GHz rotating field, at different layers of the sample, respectively.

To generate twisted spin-waves with topological charges $l=+2$ or $l=-2$, a counterclockwise or clockwise rotating field is exerted on the bottom layer of the sample with initial MV state having $p=+1$ or $p=-1$, respectively. The rotating field is spatially non-uniform and its orientation $\vec{e_p}$ at $t=0$ is schematically shown in Fig. 1(b). Figs. 4(e) and 4(f) illustrate snapshots of twisted spin-waves carrying topological charges $l=-2$ and $l=+2$, respectively, 300 ps after application of the rotating field with frequency of 7.4 GHz and amplitude of 5 mT. Two movies are included in supplemental materials I and II to display the continuous oscillation of the twisted spin-waves with $|l|=1$ and $|l|=2$, respectively.

## 5. Analytical calculation of the dispersion relation

In the following, the dispersion relationship of twisted spin-waves is derived analytically. The magnetization dynamics is described by the Landau-Lifshitz-Gilbert (LLG) equation

$$\frac{\partial \boldsymbol{m}}{\partial t} = -\gamma \mu_0 \boldsymbol{m} \times \boldsymbol{H}_{eff} + \alpha \boldsymbol{m} \times \frac{\partial \boldsymbol{m}}{\partial t}, \tag{1}$$

in which with $\gamma = 1.76 \times 10^{11} \text{rad}\,\text{s}^{-1}\,\text{T}^{-1}$ is the gyromagnetic ratio, $\mu_0$ is the vacuum permeability, and $\alpha$ is the Gilbert damping. Since $\alpha$ is small for permalloy, the second term on the right side of LLG equation is neglected and we get the Landau-Lifshitz equation

$$\partial \boldsymbol{m}/\partial t = -\gamma \mu_0 \boldsymbol{m} \times \boldsymbol{H}_{eff} \tag{2}$$

Using cylindrical coordinates $(\rho, \phi, z)$, magnetization of the initial MV state is denoted as

$$\boldsymbol{m_0} = (0,1,0) \tag{3}$$

The twisted spin-wave with topological charge $l$ can be expressed approximately as[12,22],

$$\boldsymbol{m} = (m_\rho J_l(k_\perp \rho) e^{i(l\varphi - kz - \omega t)}, 1, m_z J_l(k_\perp \rho) e^{i(l\varphi - kz - \omega t)}) \tag{4}$$

where $J_l(x)$ is the Bessel function of the first kind with order $l$. To simplify calculation of the effective field $\boldsymbol{H}_{eff}$, only the contribution of exchange field is considered. So, we have

$$\boldsymbol{H}_{eff} = A^* \nabla^2 \boldsymbol{m} \tag{5}$$

where $A^* = 2\gamma A_{ex}/M$. Substituting (4) and (5) into (2), the following equations are obtained

$$\begin{pmatrix} -i\omega m_\rho J_1(k_\perp \rho) exp(i(l\varphi - kz - \omega t)) \\ 0 \\ -i\omega m_z J_1(k_\perp \rho) exp(i(l\varphi - kz - \omega t)) \end{pmatrix}$$

$$= A^* \begin{pmatrix} -m_z(k_\perp^2 J_l + J_l k_z^2) exp(i(l\varphi - kz - \omega t)) \\ 0 \\ m_\rho(k_\perp^2 J_l + J_l k_z^2) exp(i(l\varphi - kz - \omega t)) \end{pmatrix} \tag{6}$$

By asking the coefficient determinant of equations (6) equals zero, we get the dispersion relationship:

$$\omega(k_z) = A^*(k_z^2 + k_\perp^2) \qquad (7)$$

where $k_z = 2\pi/\lambda$ and $k_\perp = x_0/r$. And $\lambda$ is wavelength of the twisted spin-wave, $x_0$ is the first zero point of the $l^{th}$-order Bessel function, and $r$ is radius of the nanodisk.

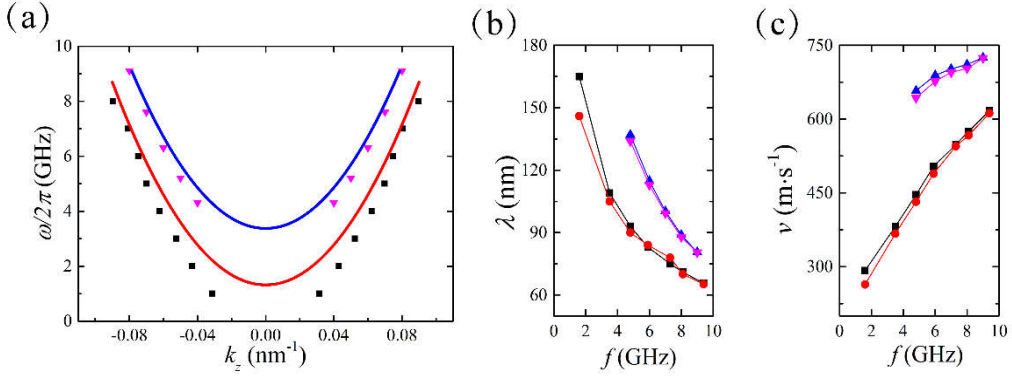

Fig.5. (a) The dispersion relationships of twisted spin-waves. The simulated and analytical results of the waves with topological charge $|l| = 1$ is represented by black squares and a solid red line, respectively. The numerical and analytical results of the waves with topological charge $|l| = 2$ is denoted by pink triangles and a solid blue line, respectively. (b) and (c) show variation of the wavelength $\lambda$ and wave speed $v$ with frequency $f$, respectively. The red dots and black squares represent the numerical and analytical results of waves with topological charge $|l| = 1$, respectively. The pink and blue triangles denote the numerical and analytical results of waves with topological charge $|l| = 2$, respectively.

Figure 5(a) shows that results of Eq. (7) agree very well with the micromagnetic numerical calculations for twisted spin-waves with frequency higher than 5 GHz. This can be attributed to the dominance of exchange energy in this frequency region, and neglecting dipolar interaction in the analytical derivation is reasonable[29]. As the frequency reduces, the discrepancy between theoretical and numerical results increases, due to the growing importance of magnetic dipole interaction.

The numerical calculations are in good agreement with the theoretical results on the variation of wavelength $\lambda$ and propagation speed $v$ with frequency $f$, as shown in Figs. 5(b) and 5(c), respectively. For twisted spin-waves with $|l| = 1$, the wavelength decreases monotonically from 146 nm to 65 nm when the frequency increases from 1.6 GHz to 9.4 GHz. The wavelength of spin-waves with $|l| = 2$ is 134 nm at $f = 4.8$ GHz, and reduces steadily to 80 nm at $f = 9$ GHz. Speed of the $|l| = 1$ twisted spin-waves is 264 m/s at $f$ = 1.6 GHz, and it increases monotonically to 612 m/s at $f$ = 9.4 GHz. The $|l| = 2$ vortex waves are much faster. It propagates 643 m/s at $f$ = 4.8 GHz and the speed increases steadily to 724 m/s at $f$ = 9 GHz.

To conclude, twisted spin-wave modes are excited in a thick permalloy nanodisk by exerting pulsed field at the bottom layer of the sample. The sign of the topological charge, carried by twisted spin-waves, depends on the core polarity of the vortex state in the sample. We then use rotating magnetic field applied at the bottom of the thick nanodisk to generate continuous oscillation of twisted spin-waves with topological charges $|l| = 1$ and $|l| = 2$. Experimentally, the rotating magnetic field can be readily generated at the cross of two perpendicular current wires. The dispersion relationship of the twisted spin-waves is calculated analytically. And the theoretical results are in good agreement with numerical calculations, especially for wave frequency higher than 5 GHz.

We acknowledge the financial support from the National Natural Science Foundation of China under grant number 11174238.